\documentclass[11pt, a4paper, logo, copyright, nonumbering]{aigreport}

\usepackage[numbers,sort&compress]{natbib}
\usepackage{xurl}
\usepackage{hyperref}
\usepackage{amsmath}
\usepackage{amssymb}
\usepackage{booktabs}
\usepackage{graphicx}
\usepackage{tabularx}
\usepackage{xcolor}
\usepackage{tikz}
\usepackage{pgfplots}
\usepackage[most]{tcolorbox}
\usepackage[capitalize,noabbrev]{cleveref}
\usetikzlibrary{arrows.meta,backgrounds,fit,positioning,shapes.geometric}
\pgfplotsset{compat=1.18}

\reportnumber{}

\definecolor{cardbg}{RGB}{226,240,255}
\definecolor{accent}{RGB}{20,110,245}
\definecolor{linkblue}{RGB}{20,110,245}
\definecolor{riskred}{RGB}{190,55,55}
\definecolor{warmorange}{RGB}{226,132,45}
\definecolor{safegreen}{RGB}{45,135,93}
\definecolor{mutedgray}{RGB}{100,112,126}

\newcommand{\paperauthors}{Xiangfan Wu, Zonghao Ying, Huiyu Wu, Xing Zheng, Huangsheng Cheng, Xiaorong Shi, Jing Guo}

\title{Ventor-QTest: Threat-Model-Driven Verification of Vendor-Hosted LLM APIs}
\author{\paperauthors}
\hypersetup{
  hidelinks,
  pdftitle={Ventor-QTest: Threat-Model-Driven Verification of Vendor-Hosted LLM APIs},
  pdfauthor={Xiangfan Wu, Zonghao Ying, Huiyu Wu, Xing Zheng, Huangsheng Cheng, Xiaorong Shi, Jing Guo}
}

\begin{abstract}
As large language models become increasingly widespread, third-party providers that deploy open-weight models have become an important part of the ecosystem.  Auditing the quality of their inference APIs is therefore an open problem.  We formalize hosted model routing as a stochastic process and propose \mbox{\textbf{Ventor-QTest}}, a composite black-box audit that requires no probability information from the target API.  Its repeated-request component sends each frozen constrained context to the target multiple times, reconstructs a categorical output distribution from the returned text counts, and reports \emph{average fidelity loss} (AFL) as a null-bias-corrected, within-window mean coarsened-KL statistic.  Its long-sequence component uses independent runs to report \emph{extreme fidelity loss} (EFL) through the empirical upper tail of a run-level reference-centered-surprisal statistic.  Across three logprob-capable route conditions, AFL shows strong linear descriptive agreement with a logprob-derived coarsened-KL comparator.  Across seven route snapshots, 20-run sequence probes reveal route-specific EFL variation.  AFL and EFL have little detectable route-level association with GPQA-Diamond accuracy.  In contrast, pronounced EFL coincides with a decline in Terminal-Bench pass rate as task exposure increases.  This pattern may arise because correctness in long-horizon tasks is more sensitive to extreme fidelity loss.  These results motivate reporting AFL and EFL jointly, particularly when auditing long-horizon agentic tasks.  The open-source implementation is available at \url{https://github.com/Tencent/AI-Infra-Guard/tree/main/services/api_checker/ventor_qtest}.

\end{abstract}

\begin{document}

\thispagestyle{firststyle}
\setlength{\parindent}{0pt}

{\LARGE\bfseries
\textcolor{accent}{Ventor-QTest}: Threat-Model-Driven Verification\\
of Vendor-Hosted LLM APIs\par}

\vskip 12pt

{\large\bfseries
Tencent Zhuque Lab\par}

\vskip 8pt

{\small
Xiangfan Wu, Zonghao Ying, Huiyu Wu, Xing Zheng,\\
Huangsheng Cheng, Xiaorong Shi, Jing Guo\par}

\begin{tcolorbox}[
  enhanced, boxrule=0pt, frame hidden,
  colback=cardbg, arc=12pt,
  left=18pt, right=18pt, top=10pt, bottom=10pt,
  before skip=12pt, after skip=14pt
]
{\bfseries\large Abstract\par}
\vskip 5pt
{\small As large language models become increasingly widespread, third-party providers that deploy open-weight models have become an important part of the ecosystem.  Auditing the quality of their inference APIs is therefore an open problem.  We formalize hosted model routing as a stochastic process and propose \mbox{\textbf{Ventor-QTest}}, a composite black-box audit that requires no probability information from the target API.  Its repeated-request component sends each frozen constrained context to the target multiple times, reconstructs a categorical output distribution from the returned text counts, and reports \emph{average fidelity loss} (AFL) as a null-bias-corrected, within-window mean coarsened-KL statistic.  Its long-sequence component uses independent runs to report \emph{extreme fidelity loss} (EFL) through the empirical upper tail of a run-level reference-centered-surprisal statistic.  Across three logprob-capable route conditions, AFL shows strong linear descriptive agreement with a logprob-derived coarsened-KL comparator.  Across seven route snapshots, 20-run sequence probes reveal route-specific EFL variation.  AFL and EFL have little detectable route-level association with GPQA-Diamond accuracy.  In contrast, pronounced EFL coincides with a decline in Terminal-Bench pass rate as task exposure increases.  This pattern may arise because correctness in long-horizon tasks is more sensitive to extreme fidelity loss.  These results motivate reporting AFL and EFL jointly, particularly when auditing long-horizon agentic tasks.  The open-source implementation is available at \url{https://github.com/Tencent/AI-Infra-Guard/tree/main/services/api_checker/ventor_qtest}.
\par}
\end{tcolorbox}

\section{Introduction}
\label{sec:introduction}

Frontier large language models (LLMs) are reshaping software development, information retrieval, and data analysis, and they increasingly serve as core reasoning components in real-world workflows.  Agentic coding systems make this shift concrete: Codex and Claude Code can inspect codebases, edit files, invoke development tools, run tests, and execute multi-step tasks~\cite{openai_codex_2026,anthropic_claude_code_2026}.  As these systems move beyond text generation to act on external state, the behavior of the model serving each request becomes a systems-level reliability concern.

Open-weight releases have simultaneously become a major part of the LLM ecosystem.  The number, adoption, and downstream reuse of public models on platforms such as Hugging Face continue to grow, sustaining an active ecosystem of base models, quantized variants, adapters, and application-specific derivatives~\cite{huggingface_state_open_source_2026,longpre2025economies}.  Open weights, however, do not imply inexpensive self-hosting.  For example, the official Kimi K2 deployment guide specifies that the smallest deployment unit for its FP8 weights at a 128K context length on H200 or H20 hardware comprises 16 GPUs~\cite{moonshot_kimi_k2_deployment_2025}.  Such infrastructure requirements encourage developers and organizations to access open models through third-party inference APIs, moving execution outside the model publisher's trusted environment.

This shift creates a verification gap.  Model names in hosted LLM APIs are claims rather than cryptographic attestations in the sense of an evidence-producing remote-attestation architecture~\cite{birkholz2023rats}.  A client who requests a particular checkpoint normally observes only a response string and provider-controlled metadata exposed by the serving APIs~\cite{deepseek2026chatapi,siliconflow2026chatapi,openrouter2026routing}.  The serving path may instead use an older checkpoint, a cheaper substitute, a quantized deployment, or a different decoding stack.  Quantization can affect inference efficiency and task accuracy~\cite{zeng2024abqllm,husom2025sustainable,wang2025fp4}; separately, deployment-verification studies show that serving configurations can alter observable output behavior~\cite{gao2025model,zhu2026rut,richardeau2026flips,cai2025substitution}.  Some changes are benign engineering choices, while others may violate an advertised contract, and black-box observations alone do not establish which explanation is true.  We therefore ask: \emph{Can black-box API queries measure the behavioral deviation of a third-party deployment from a trusted reference?}  This paper studies this question as \emph{deployment verification}: testing whether a hosted endpoint behaves like a trusted reference deployment under a stated configuration.

We instantiate this view in \textbf{Ventor-QTest}, a composite black-box audit requiring probabilities only from a trusted reference.  The repeated-request component sends each frozen constrained context to the target $M$ times, maps the returned texts into predeclared categories, reconstructs the corresponding target output distribution, and compares it with the trusted reference.  Its output $S_r$ is the \emph{average fidelity loss} (AFL): a null-bias-corrected, within-window mean coarsened-KL statistic.  Independent long-sequence probes retain the empirical distribution of all observed run-level centered-surprisal deviations; their empirical upper-tail behavior is the \emph{extreme fidelity loss} (EFL).  AFL and EFL have different units and are reported together, not collapsed by a manually calibrated threshold or weighted sum.

Experiments across hosted DeepSeek route snapshots show descriptive agreement between AFL and a logprob-derived comparator, and reveal route-specific EFL.  AFL and EFL have little detectable route-level association with GPQA-Diamond accuracy~\cite{rein2024gpqa}.  In contrast, pronounced EFL coincides with a decline in Terminal-Bench pass rate as task exposure increases~\cite{merrill2026terminalbench}.  This pattern may arise because correctness in long-horizon tasks is more sensitive to extreme fidelity loss.

This paper makes four contributions:
\begin{itemize}
    \item We formalize vendor-hosted inference as a route-level stochastic process and distinguish persistent, intermittent, substitution, and adaptive-routing deviations.
    \item We develop a text-only repeated-request estimator of AFL for a predeclared finite outcome map and demonstrate strong linear descriptive agreement with a logprob-derived coarsened-KL comparator on three logprob-capable route conditions.
    \item We introduce a complementary long-sequence probe of EFL and report the empirical distribution of all observed run-level centered-surprisal deviations, including upper-tail summaries.
    \item We evaluate AFL and EFL across hosted route snapshots and contrast their downstream relationships on GPQA-Diamond and Terminal-Bench.
\end{itemize}

\section{Scope, Threat Model, and Assumptions}
\label{sec:threat}

\subsection{Verification Target}

The object under test is a \emph{served model instance}, not weights in isolation.  We represent an advertised claim as
\begin{equation}
    C=(m, v, q, \phi, \mathcal{E}),
    \label{eq:claim}
\end{equation}
where $m$ is the model family, $v$ is the checkpoint or version, $q$ is the numerical precision or quantization, $\phi$ denotes decoding and serving parameters, and $\mathcal{E}$ denotes API semantics such as tool-call formatting.  A test can reject behavioral consistency with $C$ on a probe distribution.  It cannot, from text alone, prove which component changed or whether the change was intentional.

The claim in \cref{eq:claim} describes the full advertised service, but the present Ventor-QTest instantiation tests only its text-generation projection.  It does not validate the full API semantics $\mathcal E$, including tool execution, multimodal preprocessing, or agent-environment interaction.

The verifier controls a client and has access to a trusted reference endpoint $P$ for the claimed instance.  A static target would expose one endpoint distribution $Q_r$, but a deployed route can switch replicas, numerical formats, inference kernels, or routing policies across serving windows.  We therefore model route $r$ as a stochastic process $\{Q_{r,b}\}_{b\in\mathcal B}$, where $b$ indexes an independently sampled audit window.  Given challenges $x\sim\mathcal{G}$, full stochastic consistency is
\begin{equation}
    H_0: Q_{r,b}(\cdot\mid x,\phi)=P(\cdot\mid x,\phi)
    \quad\text{for all }x\in\mathcal{G}\text{ and }b\in\mathcal B.
    \label{eq:null}
\end{equation}
Equation~\eqref{eq:null} is the ideal global consistency condition; its violation defines a served-distribution deviation, not a statement that the target is worse on downstream tasks.  The finite audit below does not test this universal condition directly.

This process view separates persistent and intermittent latent divergence.  Let $T_x$ be the declared outcome coarsening for challenge $x$ and define the latent window divergence
\begin{equation}
 L_{r,b}=\mathbb E_{x\sim\mathcal G}
 D_{\mathrm{KL}}\!\left(T_xQ_{r,b}\,\middle\|\,T_xP\right),
 \qquad A_r=\mathbb E_b[L_{r,b}].
 \label{eq:window-loss}
\end{equation}
The cross-window mean $A_r$ describes a persistent shift.  For a task exposed to $m$ independently sampled serving windows, define the latent excess-maximum functional
\begin{equation}
 X_r(m)=\mathbb E\!\left[\max_{1\le k\le m}L_{r,k}\right]-A_r.
 \label{eq:extreme-loss}
\end{equation}
Unlike a thresholded degraded/not-degraded label, $X_r(m)$ retains the magnitude of the latent upper tail and is nondecreasing in the number of sampled windows $m$.  A route can therefore have a modest $A_r$ but a material $X_r(m)$ when severe deviations occur intermittently.  This makes explicit a stochastic serving-window dimension that is implicit in route-level capability verification such as KVV~\cite{moonshot2025k2vv,moonshot2026kvvblog}.  The primary experiment does not directly estimate either $A_r$ or $X_r(m)$; they motivate the two observable audit outputs summarized in \cref{tab:latent-observed}.

\begin{table}[ht]
\centering
\caption{Latent objects and observed Ventor-QTest statistics.  AFL and EFL name the two reported observables; they are not interchangeable estimators of a common loss.}
\label{tab:latent-observed}
\small
\begin{tabularx}{\textwidth}{@{}p{0.16\textwidth}>{\raggedright\arraybackslash}X >{\raggedright\arraybackslash}p{0.25\textwidth}@{}}
\toprule
Object & Meaning & Directly estimated here? \\
\midrule
$A_r$ & Cross-window mean latent coarsened KL & No; most routes have one repeated-request window \\
$S_{r,b}$ & AFL: within-window, across-context bias-corrected coarsened-KL statistic & Yes \\
$X_r(m)$ & Expected excess maximum of latent window KL & No \\
$D_{r,b}$, $\widehat F_r^D$ & EFL: run-level centered-surprisal statistic and its observed empirical upper tail & Yes; neither is KL \\
\bottomrule
\end{tabularx}
\end{table}

For tasks whose desired behavior is defined by the trusted reference, greater model-output exposure may create more opportunities to encounter an altered serving window.  We treat this statement as motivation for exploratory downstream analysis rather than a deterministic task-success model.  Deviations may hurt, leave unchanged, or occasionally improve a task outcome, while timeouts, rate limits, and tool semantics provide additional non-model pathways to failure.

\subsection{Provider Behavior Classes}

We consider five increasingly adversarial cases.
\begin{description}
    \item[Operational deviation.] An honest vendor uses the claimed weights but differs in inference engine, parameter validation, tokenization, batching, or numerical kernels.
    \item[Undisclosed quantization.] The vendor serves a lower-precision instance, such as FP8 or FP4, without making the distinction clear to the client.  FP8 is an established low-precision format, whereas FP4 is an actively studied lower-precision regime~\cite{micikevicius2022fp8,wang2025fp4}; deployment verification studies show that quantization can change observable model behavior without implying malicious intent~\cite{gao2025model,zhu2026rut,richardeau2026flips}.
    \item[Intermittent fidelity degradation.] A heterogeneous replica pool or time-varying router usually serves a close configuration but occasionally selects a substantially altered one.  This case may leave a moderate average deviation while enlarging the upper tail across independent runs.
    \item[Substitution or metadata forgery.] The vendor serves another checkpoint or model and may return the requested model name in provider-controlled metadata.  We therefore treat response metadata as evidence to preserve, not as ground truth.
    \item[Adaptive selective routing.] The vendor observes the request, classifies whether it is an audit, and routes likely probes to a faithful model while routing other traffic to a cheaper or altered model.  It may maintain state across requests, cache known answers, or fine-tune a substitute to imitate public probes.  Recent work studies adversarial substitution and diluted or mixed routing~\cite{cai2025substitution,fang2026kbf,zhang2026iris} and demonstrates that finite-budget fingerprints can be spoofed by a model trained to mimic their probe behavior~\cite{zhang2026ghostprint}.
\end{description}

The adaptive case exposes the limitation of a fixed question set.  Let $h(x)\in\{0,1\}$ be the provider's probe classifier and let $R$ be a substitute.  Selective routing implements
\begin{equation}
 Q_{\mathrm{route}}(\cdot\mid x)=
 \begin{cases}
 P(\cdot\mid x), & h(x)=1,\\
 R(\cdot\mid x), & h(x)=0.
 \end{cases}
 \label{eq:routing}
\end{equation}
For a public finite set $\mathcal{S}$, the dictionary classifier $h(x)=\mathbb{1}[x\in\mathcal{S}]$ defeats any audit restricted to $\mathcal{S}$.  A hidden set delays this attack but does not remove it: leakage, repeated observation, or surrogate reconstruction can turn the hidden set into a known one.  Online selective routing can therefore defeat audits based on recognizable finite probes~\cite{pasquini2025llmmap,cai2025substitution,fang2026kbf,zhang2026iris,zhang2026ghostprint}.  Separately, benchmark contamination undermines inference from performance on known tests to behavior on unseen traffic~\cite{white2025livebench,zhu2024decontamination}.

\subsection{Assumptions and Non-goals}
\label{sec:assumptions}

Our conclusions rely on the following assumptions.
\begin{enumerate}
    \item[\textbf{A1}.] The reference endpoint is authentic and is not colluding with the tested vendor.
    \item[\textbf{A2}.] Reference and target requests use the same declared sampling configuration, and gross parameter rejection is checked before behavioral comparison.
    \item[\textbf{A3}.] The reference supports next-token log probabilities and assistant-prefix continuation, as specified by the reference API~\cite{deepseek2026chatapi,deepseek2026prefix}.  This capability is required only from the reference: without its conditional probabilities, the surprisal, entropy, and information variance are undefined.  The target requires only ordinary text generation.  A tail run is admitted only when the exact target prefix is accepted by the reference and all 500 prespecified positions receive valid conditional probabilities.  This is an API-level replay-and-scoring gate, not evidence that the two routes use identical tokenizers.
    \item[\textbf{A4a}.] The serving distribution is approximately stationary within one audit window.  It may change across independently collected windows; this between-window variation is part of the process model rather than noise to discard.
    \item[\textbf{A4b}.] Conditional on the declared request configuration and audit window, repeated requests within a context--route cell are treated as approximately independent draws from a fixed route distribution.  Cache affinity, batching, or session routing can violate this approximation and produce overdispersion.
    \item[\textbf{A5}.] The collected reference category vector is treated as fixed for inference.  Sampling uncertainty in that API-returned vector and reference drift across serving windows are not propagated into the reported intervals or null calibration.
    \item[\textbf{A6}.] A runtime nonce is secret before a challenge is sent, but the provider can inspect the challenge on arrival.  Freshness therefore prevents exact precomputation, not semantic probe detection.
\end{enumerate}

Ventor-QTest is not a proof of model ownership, remote attestation, or malicious intent.  It does not identify a quantization level from a deviation score, certify downstream quality, or guarantee detection when a provider can perfectly distinguish audit traffic.  The audit complements capability tests such as KVV: a deployment can be distributionally close on a short text-generation probe yet fail tool calling, multimodal preprocessing, long-context inference, or agentic execution.

\section{Method}
\label{sec:method}

\subsection{Overview}

\begin{figure}[t]
\centering
\resizebox{0.98\textwidth}{!}{%
\begin{tikzpicture}[
    font=\small,
    >=Latex,
    flow/.style={-Latex,line width=0.9pt,draw=mutedgray},
    module/.style={draw=accent,fill=accent!7,rounded corners=3pt,
        minimum height=1.55cm,text width=3.35cm,align=center,line width=0.9pt,
        inner sep=5pt},
    target/.style={draw=warmorange,fill=warmorange!9,rounded corners=3pt,
        minimum height=1.55cm,text width=3.35cm,align=center,line width=0.9pt,
        inner sep=5pt},
    reference/.style={draw=safegreen,fill=safegreen!8,rounded corners=3pt,
        minimum height=1.55cm,text width=3.35cm,align=center,line width=0.9pt,
        inner sep=5pt}
]

\node[module] (contexts) at (0,1.50) {\textbf{Repeated requests}\\$M$ calls for each fixed $x_j$};
\node[target] (target) at (4.55,2.45) {\textbf{Target text counts}\\$N_{jri}$ from $M$ samples\\per context};
\node[reference] (reference) at (4.55,0.55) {\textbf{Trusted reference}\\category vectors\\$\pi_1,\ldots,\pi_J$};
\node[module] (estimate) at (9.10,1.50) {\textbf{Bias-corrected KL}\\per-context $\widehat K^{\mathrm{bc}}_{jr}$\\and aggregation};
\node[module] (mean) at (13.65,1.50) {\textbf{Average fidelity loss}\\(AFL) $S_r$ with\\joint-null inference};

\node[module] (fresh) at (0,-1.90) {\textbf{Fresh sequence runs}\\$B$ independent nonce\\challenges};
\node[target] (sequence) at (4.55,-1.90) {\textbf{Complete target runs}\\$Y_{rb,1:T}$\\all positions retained};
\node[reference] (rescore) at (9.10,-1.90) {\textbf{Reference rescoring}\\centered surprise\\$e_{rbt}$};
\node[module] (tail) at (13.65,-1.90) {\textbf{Extreme fidelity loss}\\(EFL) empirical tail\\of $\widehat F_r^D$};

\draw[flow] (contexts.east) to[out=25,in=180] (target.west);
\draw[flow] (contexts.east) to[out=-25,in=180] (reference.west);
\draw[flow] (target.east) to[out=0,in=180] ([yshift=4mm]estimate.west);
\draw[flow] (reference.east) to[out=0,in=180] ([yshift=-4mm]estimate.west);
\draw[flow] (estimate.east) -- (mean.west);
\draw[flow] (fresh.east) -- (sequence.west);
\draw[flow] (sequence.east) -- (rescore.west);
\draw[flow] (rescore.east) -- (tail.west);

\end{tikzpicture}%
}
\caption{Composite Ventor-QTest pipeline.  Repeated requests to fixed contexts report AFL as a within-window mean coarsened-KL statistic, while independent long-sequence runs report EFL from the empirical upper tail of all observed centered-surprisal deviations.  Both components require probabilities only from the trusted reference.  Their outputs have different units and are reported jointly rather than summed into a scalar.}
\label{fig:pipeline}
\end{figure}
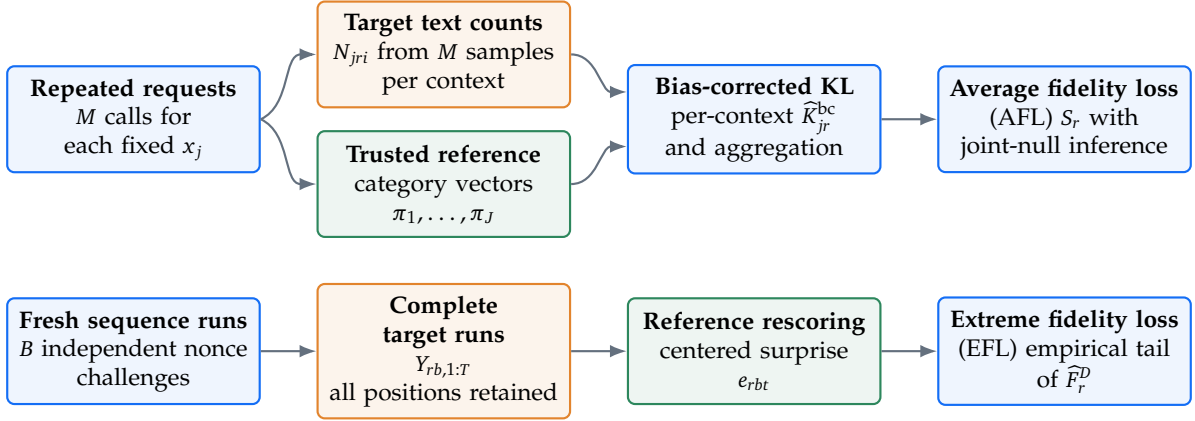

Ventor-QTest is a composite black-box audit with complementary AFL and EFL components.  As shown in \cref{fig:pipeline}, the repeated-request AFL component reconstructs finite conditional output distributions from repeated target text and returns a bias-corrected, within-window coarsened-KL statistic.  The long-sequence EFL component repeats independently generated probes and retains the empirical distribution of all observed run-level deviations instead of collapsing it to an average or binary state.  These observables have different units and are reported as a pair; they are never added into an arbitrarily weighted scalar.  The procedure is specified in \cref{app:procedure}.

\subsection{Repeated-request Outcome Model}
\label{sec:repeated-kl}

For each context $x_j$, a predeclared total map $g_j$ assigns every possible returned string to a finite alphabet $\mathcal A_j$.  The trusted reference induces category probabilities
\begin{equation}
 \pi_{ji}=\Pr_P[g_j(Y)=i\mid x_j],
 \qquad i\in\mathcal A_j.
 \label{eq:reference-categories}
\end{equation}
The verifier then sends the exact same context to target route $r$ a total of $M$ times.  Under the conditional-independence approximation in A4b, its category counts satisfy
\begin{equation}
 (N_{jr1},\ldots,N_{jr|\mathcal A_j|})
 \sim\operatorname{Multinomial}(M,\theta_{jr}),
 \qquad
 \theta_{jri}=\Pr_{Q_r}[g_j(Y)=i\mid x_j].
 \label{eq:repeated-multinomial}
\end{equation}
This sampling design identifies $\theta_{jr}$ from ordinary text.  Whitespace-only, multi-token, missing, or otherwise nonconforming responses map to \texttt{OTHER}, so every request contributes to the likelihood.

Low-reference-mass categories are pooled before target sampling.  Any declared category satisfying $M\pi_{ji}<c$ is merged into \texttt{OTHER}; the primary protocol fixes $c=1$.  If $T_j$ denotes this deterministic coarsening, the data-processing inequality gives
\begin{equation}
 D_{\mathrm{KL}}(T_jQ_r\|T_jP)
 \le D_{\mathrm{KL}}(Q_r\|P).
 \label{eq:data-processing}
\end{equation}
The estimand is therefore a conservative task-coarsened divergence with a fully specified outcome map, by the data-processing inequality~\cite{cover2006elements}.  Because pooling reads only $\pi_j$, the target sample cannot influence category selection.

\subsection{Finite-sample KL Estimator}
\label{sec:kl-estimator}

A reference-centered Dirichlet posterior stabilizes multinomial KL estimation while preserving the reference distribution as the prior mean.  We set
\begin{equation}
 \theta_{jr}\sim\operatorname{Dirichlet}(\tau\pi_j),
 \qquad
 a_{jri}=N_{jri}+\tau\pi_{ji},
 \qquad A_0=M+\tau,
 \label{eq:reference-prior}
\end{equation}
with total prior strength $\tau=1$.  The posterior expected coarse KL is
\begin{equation}
 \widehat K^{\mathrm{post}}_{jr}
 =\sum_i\frac{a_{jri}}{A_0}
 \left[\psi(a_{jri}+1)-\psi(A_0+1)-\log\pi_{ji}\right],
 \label{eq:posterior-kl}
\end{equation}
where $\psi$ is the digamma function.  The expression follows from the standard Dirichlet moment for $\mathbb E[\theta_i\log\theta_i]$~\cite{wolpert1995estimating}.

Finite-sample posterior uncertainty produces a positive baseline even under the null.  We estimate and subtract that baseline separately for each context:
\begin{align}
 b_j(P,M)
 &=\mathbb E_{N^{(0)}\sim\operatorname{Multinomial}(M,\pi_j)}
   [\widehat K^{\mathrm{post}}(N^{(0)},\pi_j)],\\
 \widehat K^{\mathrm{bc}}_{jr}
 &=\widehat K^{\mathrm{post}}_{jr}-b_j(P,M), &
 S_r&=\frac{1}{J}\sum_{j=1}^{J}\widehat K^{\mathrm{bc}}_{jr}.
 \label{eq:bias-corrected-kl}
\end{align}
The implementation uses 20,000 parametric-null draws to estimate $b_j$.  We refer to $S_r$ as the route's AFL in the frozen audit window.  Negative finite-sample values are retained because clipping would bias correlations and route tests upward.  A prior-free Pearson-divergence U-statistic, defined in \cref{app:repeated-context-protocol}, checks whether the empirical ordering depends on the Dirichlet correction.

When a window index is needed, we write $S_{r,b}$ for the value of $S_r$ computed separately in window $b$.  For $B$ independently spaced repeated-request windows, $B^{-1}\sum_b S_{r,b}$ would be a sample analogue of $A_r$ under the stated reference-stability and sampling assumptions, where each $S_{r,b}$ is calculated before pooling observations across windows.  The primary matched-route collection below has one window and therefore reports only its within-window context average; it does not directly estimate the cross-window quantity $A_r$.

\subsection{Auxiliary Comparator and Route Inference}

When a route exposes top log probabilities, its alternatives are aggregated through the same $g_j$ to obtain a logprob-derived coarsened-KL comparator.  This provider-controlled field is an auxiliary consistency check rather than ground truth and never enters the text-count statistic.  Association is measured after centring both quantities within context.  Exact significance permutes a common route-condition label assignment across all contexts, preserving the crossed repeated-measures structure rather than treating context--route cells as independent routes.

The implemented mean-component test concerns the finite probe null
\begin{equation}
 H^{\mathrm{mean}}_{0,r}:\quad \theta_{jr}=\pi_j,
 \qquad j=1,\ldots,J.
 \label{eq:finite-probe-null}
\end{equation}
The statistic $S_r$ is compared with a joint null generated by independent multinomial draws from the fixed $\pi_j$ across all $J$ contexts.  One-sided $p$ values use 20,000 draws, and the Holm procedure controls family-wise error across the matched routes~\cite{holm1979simple}.  Rejecting \cref{eq:finite-probe-null} establishes inconsistency only for the frozen contexts, coarsened outcome maps, declared request configuration, and collection window.  It does not directly reject the universal consistency condition in \cref{eq:null}.  The request counts are
\begin{equation}
 N_{\mathrm{target}}=RJM,
 \qquad N_{\mathrm{reference}}=J,
 \label{eq:cost}
\end{equation}
because each reference vector is shared by all $R$ routes.

\subsection{Run-Level Tail Probe}
\label{sec:tail-probe}

The run-level component preserves deviations that a route average can hide.  In independent run $b$, target route $r$ first generates a length-$T$ sequence $Y_{rb,1:T}$.  At position $t$, the trusted reference supplies the conditional distribution $p_{rbt}(\cdot)=P(\cdot\mid Y_{rb,<t})$.  Using the standard information-theoretic notions of surprise, entropy, and information variance~\cite{shannon1948mathematical,kontoyiannis2014lossless}, we calculate the centered reference surprise
\begin{equation}
 e_{rbt}=-\log p_{rbt}(Y_{rb,t})
 -H(p_{rbt}),
 \qquad
 H(p)=-\sum_y p(y)\log p(y),
 \label{eq:centered-surprise}
\end{equation}
and the complete-run deviation
\begin{equation}
 D_{r,b}(T)=\left|\frac{1}{T}\sum_{t=1}^{T}e_{rbt}\right|.
 \label{eq:run-deviation}
\end{equation}
If the target follows the reference conditional distribution, $e_{rbt}$ has zero conditional expectation, giving a martingale-difference structure under the reference null~\cite{hall1980martingale}.  Because positions share an autoregressive history, however, the statistical unit is the complete sequence, not an individual token.  All intervals therefore resample independent runs as clusters, following the nonparametric bootstrap principle~\cite{efron1993bootstrap}.

Across $B$ runs, the empirical distribution $\widehat F_r^D$ records both typical and upper-tail stochastic deviation.  We refer to the observed upper-tail behavior of $\widehat F_r^D$ as the route's EFL and report its median and standard deviation together with an upper empirical quantile and the observed maximum.  These are descriptive summaries of the observed runs, not estimates of the latent $X_r(m)$ in \cref{eq:extreme-loss}.

The sequence statistic is not a KL estimator.  Under target distribution $Q$, its signed expectation contains $D_{\mathrm{KL}}(Q\|P)+H(Q)-H(P)$, so an unknown entropy difference can offset KL.  The composite audit therefore assigns distinct roles to its components and returns
\begin{equation}
 \mathcal V_r=\left(S_r,\widehat F_r^D\right),
 \label{eq:composite-report}
\end{equation}
where $S_r$ reports AFL over the frozen contexts and $\widehat F_r^D$ reports the empirical run-to-run distribution from which EFL is summarized.  No numerical conversion between the two statistics, or from either statistic to task success, is assumed.

\section{Evaluation}
\label{sec:evaluation}

\subsection{Evaluation Design}

The experiments evaluate three questions.  We first compare repeated-request AFL with an auxiliary logprob-derived comparator, then ask whether independent runs reveal EFL hidden by typical deviation, and finally examine their downstream relationships.  The AFL, EFL, and downstream studies cover seven routes.  Frozen protocol details and the configuration fields preserved by the collection artifacts appear in Appendix~\ref{app:repeated-context-protocol}, including Sections~\ref{app:tail-protocol} and~\ref{app:downstream-protocol}; fields not preserved are stated explicitly rather than reconstructed after the fact.

\subsection{Descriptive Agreement with a Logprob-Derived Coarsened-KL Comparator}
\label{sec:repeated-validation}

\begin{figure}[ht]
    \centering
    \includegraphics[width=\textwidth]{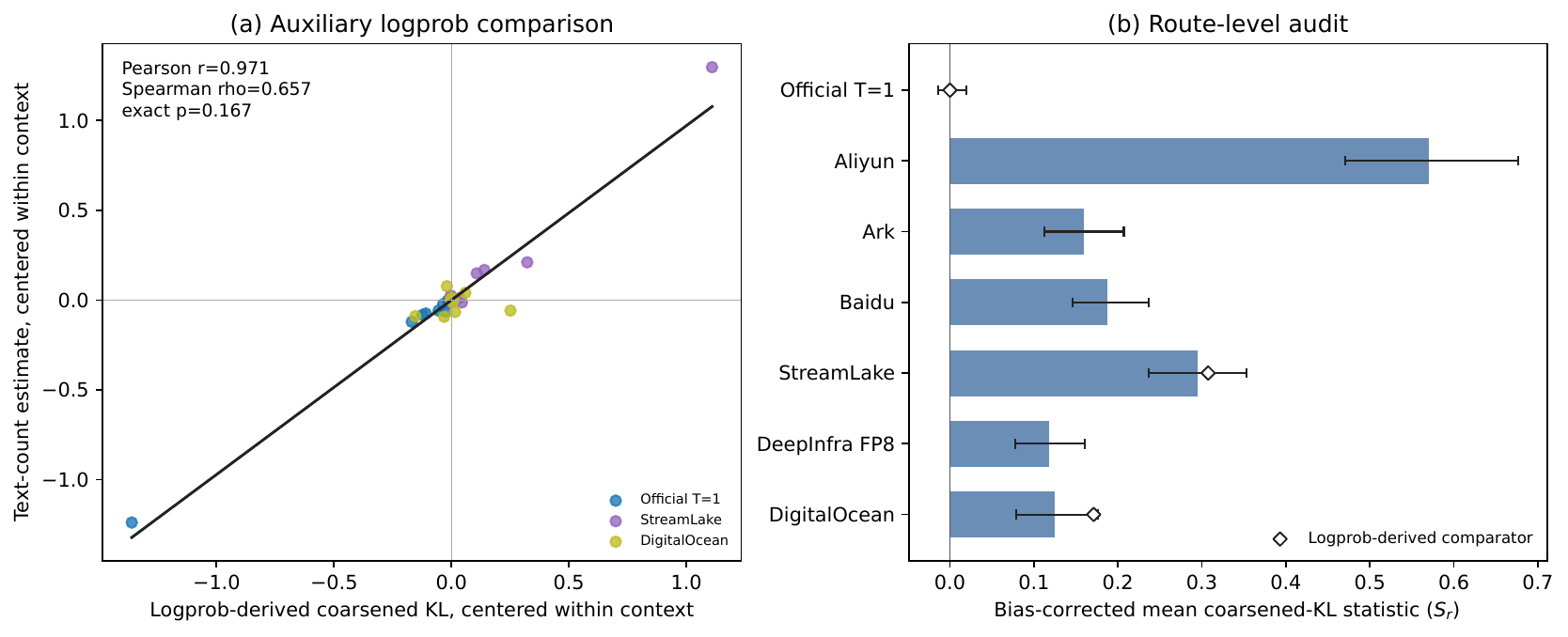}
    \caption{Auxiliary consistency check for AFL.  The left panel contains 36 context--route cells from three logprob-capable route conditions.  Text-count AFL statistics and the logprob-derived coarsened-KL comparator show descriptive agreement after context centring; with only $3!=6$ route permutations, the exact route-level test is not confirmatory.  The right panel shows the seven-route audit, in which the official control remains at the route-level null.}
    \label{fig:repeated-context-validation}
\end{figure}

The text-count AFL statistic shows strong linear descriptive agreement with the comparator derived from log probabilities.  After removing context fixed effects, Pearson correlation is $r=0.971$ and Spearman correlation is $\rho=0.657$.  Among the $3!=6$ exact common route permutations, both one-sided tests give $p=0.167$; the route-limited result is therefore descriptive rather than confirmatory.  Averaging first by route gives Pearson $r=0.989$ and perfect rank agreement.  The prior-free Pearson-divergence sensitivity and both stricter reference-only pooling thresholds retain positive association; complete values appear in \cref{app:repeated-context-protocol}.  Because the comparator is supplied by the target route, this analysis checks internal consistency rather than validating against independent ground truth.

\subsection{Matched-route AFL Audit}

\begin{table}[ht]
\centering
\caption{AFL audit over seven routes.  $S_r$ is the mean bias-corrected coarsened-KL statistic over 12 held-out contexts; intervals are 95\% posterior credible intervals after subtracting the null baseline.  The final column reports Holm-adjusted $p$ values.}
\label{tab:repeated-context-results}
\small
\begin{tabular}{lccc}
\toprule
Route & $S_r$ [95\% CrI] & Logprob comparator & Holm $p$ \\
\midrule
Official self-check & $-0.0007$ [$-0.0144$, $0.0193$] & $0.0001$ & 0.515 \\
Aliyun 0731 & $0.5704$ [$0.4710$, $0.6765$] & -- & 0.00035 \\
Ark 0731 & $0.1591$ [$0.1129$, $0.2071$] & -- & 0.00035 \\
Baidu Qianfan 0731 & $0.1875$ [$0.1456$, $0.2365$] & -- & 0.00035 \\
StreamLake & $0.2950$ [$0.2364$, $0.3533$] & $0.3075$ & 0.00035 \\
DeepInfra FP8 & $0.1185$ [$0.0782$, $0.1611$] & -- & 0.00035 \\
DigitalOcean & $0.1250$ [$0.0789$, $0.1766$] & $0.1712$ & 0.00035 \\
\bottomrule
\end{tabular}
\end{table}

The official control is centered at zero and is not rejected, while all six third-party routes reject the finite-probe joint null after multiplicity correction.  These results establish inconsistency only on the 12 frozen contexts, their coarsened outcome maps, the declared request configuration, and the collection window.  They neither reject the universal null in \cref{eq:null} nor identify a changed checkpoint, numerical precision, or serving policy.

\subsection{EFL Is Not Captured by AFL}
\label{sec:tail-results}

The long-sequence probe reveals EFL that AFL does not show.  Each entry in \cref{tab:tail-results} summarizes 20 complete 500-position runs; no run or high-deviation observation is removed.  The median characterizes typical stochastic deviation, while the standard deviation, empirical 0.95 quantile, and maximum summarize EFL in the observed runs.

\begin{table}[ht]
\centering
\caption{EFL summaries across seven route snapshots.  All quantities summarize $D_{r,b}(500)$ over 20 paired runs.  Smaller values indicate less observed absolute centered-surprise drift under this probe; the statistic is not a KL estimator and does not provide a universal ordering of model fidelity.  The empirical $q_{0.95}$ uses NumPy's linear interpolation and, with 20 runs, is descriptive.}
\label{tab:tail-results}
\small
\begin{tabular}{lrrrr}
\toprule
Route & Median $D_{500}$ & SD & Empirical $q_{0.95}$ & Maximum \\
\midrule
Official self-check & 0.00545 & 0.00835 & 0.02484 & 0.03498 \\
Aliyun 0731 & 0.00754 & 0.00819 & 0.02190 & 0.02924 \\
Ark 0731 & 0.00679 & 0.00881 & 0.02633 & 0.02901 \\
Baidu Qianfan 0731 & 0.00612 & \textbf{0.00534} & \textbf{0.01693} & \textbf{0.01975} \\
StreamLake & \textbf{0.00409} & 0.01390 & 0.04817 & 0.04969 \\
DeepInfra FP8 & 0.00517 & 0.00806 & 0.01996 & 0.03497 \\
DigitalOcean & 0.00876 & 0.01676 & 0.05414 & 0.05960 \\
\bottomrule
\end{tabular}
\end{table}

DigitalOcean is the route snapshot with the largest median, run-level standard deviation, empirical 0.95 quantile, and maximum EFL summaries in this collection.  It has the largest observed deviation in 5 of the 20 paired runs and ranks in the top two in 9 runs; these counts are descriptive.  StreamLake also has pronounced EFL despite the smallest median, illustrating the point of the composite report: AFL and EFL need not order route snapshots in the same way.

In an exploratory replication, four additional DigitalOcean repeated-request batches give
\[
 S_r\in\{0.12505,\ 0.13667,\ 0.12092,\ 0.14797\},
\]
with mean $0.13265$ and raw SD $0.01220$.  Four observations are insufficient to estimate a stable cross-window upper tail.  We therefore use the 20-run sequence study to describe the observed stochastic upper-tail phenomenon and do not claim that the four coarse-KL batches identify the latent $X_r(m)$.

\subsection{Exploratory Downstream Observations by Task Exposure}
\label{sec:downstream-results}

Across routes, AFL and EFL have little detectable association with GPQA-Diamond accuracy, and AFL alone does not order Terminal-Bench success.  In contrast, DigitalOcean---the route with the most pronounced observed EFL---shows a decline in Terminal-Bench pass rate from 82.6\% in the lowest-exposure quartile to 13.6\% in the highest-exposure quartile.  This concurrence may arise because correctness in long-horizon tasks is more sensitive to extreme fidelity loss: longer trajectories contain more model decisions and therefore more opportunities for an extreme deviation to affect the final outcome.  The comparator-relative gaps are not monotonic across the intermediate quartiles, so this mechanism remains an exploratory interpretation rather than a monotonic exposure law.  Complete analyses appear in \cref{app:downstream-results}.

\section{Related Work}
\label{sec:related}

\paragraph{Model equality and API auditing.}
Gao et al. formulate vendor verification as a two-sample model equality test and report distributional inconsistencies among commercial Llama endpoints~\cite{gao2025model}.  RUT maps a target response into a randomized rank using reference samples and tests rank uniformity under user-like prompts~\cite{zhu2026rut}.  Cai et al. formalize adversarial model substitution, evaluate text- and logprob-based auditors under production nondeterminism, and motivate hardware-backed verification~\cite{cai2025substitution}.  KBF targets relay and reseller APIs with knowledge-boundary fingerprints and evaluates low-fraction mixed routing~\cite{fang2026kbf}.  Ventor-QTest instead estimates a within-window task-coarsened conditional-KL statistic from repeated target text and reference probabilities, requiring neither target logits nor a local target-weight replica.

\paragraph{API change tracking.}
Log Probability Tracking monitors API changes with one output token and target-provided log probabilities~\cite{chauvin2026logprob}.  B3IT uses strict black-box ``border inputs'' and observes only output tokens~\cite{chauvin2026tokenefficient}; You've Changed instead compares distributions of linguistic features in generated text~\cite{dima2025changed}.  Ventor-QTest reconstructs a predeclared categorical distribution from repeated text and pairs its mean coarsened-KL statistic with an independent long-sequence empirical-tail measurement.

\paragraph{Behavioral fingerprints.}
LLMmap identifies versions with optimized queries~\cite{pasquini2025llmmap}.  FLIPS uses pseudo-random behavior to distinguish served configurations~\cite{richardeau2026flips}, while ESF optimizes sensitive token positions for black-box tamper detection~\cite{bai2025esf}.  These methods primarily address attribution, instance identity, or tamper detection.  Ventor-QTest reports a reference-relative effect on a declared finite outcome space without training a classifier.

\paragraph{Concurrent work.}
Two July 2026 preprints are concurrent with this work.  One Token Is Enough fingerprints and verifies models from empirical single-token output distributions~\cite{bruckner2026onetoken}.  IRIS audits whole-stream substitution and fractional routing dilution from text alone while sizing its own query budget~\cite{zhang2026iris}.  Both reinforce the relevance of text-only distributional probes; Ventor-QTest differs by reporting separate within-window AFL and empirical run-level EFL rather than model attribution or routing-fraction estimation.

\paragraph{Capability verification.}
Capability benchmarks complement identity tests by evaluating task behavior, including OCR~\cite{liu2024ocrbench}, multimodal reasoning~\cite{yue2025mmmupro}, software engineering~\cite{jimenez2024swebench}, mathematics~\cite{maa2026aime}, and broad knowledge~\cite{wang2024mmlupro,white2025livebench}.  Kimi Vendor Verifier (KVV) applies this view to hosted deployments: its evaluations cover tool triggering and schema correctness, multimodal processing, long-output reasoning, and agentic coding~\cite{moonshot2025k2vv,moonshot2026kvvblog}.  KVV therefore establishes the practical importance of deployment correctness and already recognizes benchmark fluctuation.  Ventor-QTest formalizes a complementary dimension: it treats a route as a distribution over serving windows and separates persistent mean deviation from intermittent upper-tail deviation before a full capability benchmark is run.

\paragraph{Query budget and monetary cost.}
\begin{table}[t]
\centering
\caption{Representative operating points for black-box verification methods.  Counts are not matched-power comparisons.  Ventor-QTest's mean component is target-side inexpensive, whereas its run-level component is reference-intensive.}
\label{tab:related-budget}
\small
\begingroup
\setlength{\tabcolsep}{3.5pt}
\renewcommand{\arraystretch}{1.15}
\begin{tabularx}{\textwidth}{@{}>{\raggedright\arraybackslash}p{0.14\textwidth}>{\raggedright\arraybackslash}p{0.18\textwidth}>{\raggedright\arraybackslash}p{0.25\textwidth}>{\raggedright\arraybackslash}X@{}}
\toprule
Method & Target calls & Reference/setup & Audit objective and limitation \\
\midrule
Gao et al.~\cite{gao2025model} & 200--250 per single test; 10 tests/verdict & Matched 200--250 reference samples per test & Distributional equality/distance; adaptive routing not explicit \\
RUT~\cite{zhu2026rut} & 100 & 10,000 local generations and logits & Rank uniformity on user-like prompts; probabilistic substitution model \\
LLMmap~\cite{pasquini2025llmmap} & 3--8 online & About 1,200 sourcing calls per reference model over train/test; trained classifier & Closed-set version identity; analyzes query-informed defenses \\
FLIPS~\cite{richardeau2026flips} & 8 & 40 extraction calls; trained classifier & Closed/open-set instance identity; assumes a non-adaptive target \\
KVV~\cite{moonshot2025k2vv,moonshot2026kvvblog} & 4,000 & Official responses and labels & Functional behavior; a finite suite can be recognized \\
\midrule
Ventor-QTest & Mean: 600 one-token; run level: 20 sequences & Mean: 12 shared probability calls; run level: 10,000 prefix-rescoring calls/route & Finite-probe mean statistic plus empirical run tail; no guarantee against perfect recognition \\
\bottomrule
\end{tabularx}
\endgroup
\end{table}

As \cref{tab:related-budget} shows, request counts purchase different forms of evidence.  Gao et al.'s 200--250 target calls denote one two-sample test, with an equal reference sample; their endpoint verdict aggregates 10 repeated tests.  LLMmap's setup is per reference model: 75 configurations times eight queries for training and a disjoint construction of the same size for testing, distinct from its 3--8-call online verification.  Fingerprinting methods minimize online verification cost by amortizing a trained reference classifier, RUT shifts most inference to a local reference, and KVV spends substantially more target tokens to measure functional capability directly.  Ventor-QTest separates a target-side inexpensive mean component from a reference-intensive run-level component.  The mean audit uses one-token target calls, shares 12 reference vectors across routes, and returns a finite-probe coarsened-KL statistic with a calibrated decision at a declared power target (\cref{app:repeated-power,app:cost-accounting}); the run-level audit spends 10,000 reference prefix-rescoring calls per route to preserve 20 complete observations.  Monetary accounting is reported only for the mean component whose exact metering is available.

\paragraph{Adaptive audit recognition.}
Public probes create a selective-routing threat.  LLMmap analyzes query-informed perturbation of recognized fingerprint traffic, RUT uses ordinary task prompts to reduce recognizability, KBF and IRIS test mixed or diluted routing, and GhostPrint demonstrates finite-budget fingerprint imitation~\cite{pasquini2025llmmap,zhu2026rut,fang2026kbf,zhang2026iris,zhang2026ghostprint}.  Ventor-QTest's contexts are therefore evidence only for the stated audit distribution; prompt secrecy is not treated as protection against a provider that recognizes the task family.

\section{Discussion and Limitations}
\label{sec:discussion}

\paragraph{Measurement scope.}
The mean component reports coarsened KL only on the declared maps and within the observed window.  The run-level component measures centered-surprise deviation rather than KL.  The present sample supports comparison of observed empirical distributions but not precise rare-event probabilities or estimation of $X_r(m)$.  Multinomial calibration treats within-cell requests as approximately independent and the returned reference vectors as fixed; correlated routing, overdispersion, reference uncertainty, or reference drift can make intervals and null $p$ values optimistic.  Like any behavioral audit, Ventor-QTest cannot detect deviations outside its probe distribution or defeat a provider that perfectly recognizes audit traffic.

\paragraph{Downstream scope.}
The paper does not posit a monotonic mapping from AFL or EFL to task success.  A deviation may hurt, leave unchanged, or improve a particular task, and benchmark outcomes also reflect timeouts, rate limits, tool semantics, and agent-runtime behavior.  The observed concurrence between pronounced EFL and decreasing Terminal-Bench pass rate suggests that extreme fidelity loss may influence correctness in long-horizon tasks, but the unsynchronized probe and benchmark windows cannot distinguish this explanation from those alternatives.  Confirming the causal contribution requires more independent windows and time-aligned probe--benchmark collection.

\section{Conclusion}
\label{sec:conclusion}

Ventor-QTest returns two observables that are not interchangeable: AFL, a null-bias-corrected within-window coarsened-KL statistic, and EFL, the empirical upper-tail behavior of run-level centered-surprisal drift.  Experiments show descriptive agreement between AFL and a logprob-derived comparator, and show that AFL and EFL can order route snapshots differently.

The downstream comparison finds little association between either audit output and GPQA-Diamond accuracy, while pronounced EFL coincides with declining Terminal-Bench pass rate as task exposure grows.  This pattern suggests that extreme fidelity loss may matter more for correctness in long-horizon tasks.  The central methodological implication is to report AFL and EFL jointly while stating the finite probe scope of each inference.

\bibliography{ventor_qtest}

\appendix

\section{Audit Procedure}
\label{app:procedure}

For fixed contexts $x_1,\ldots,x_J$, target routes $Q_1,\ldots,Q_R$, and trusted reference $P$, the implemented audit executes the following steps:
\begin{enumerate}
    \item Freeze every context, total outcome map $g_j$, repetition count $M$, pooling threshold $c$, prior strength $\tau$, multiplicity family, and analysis seed before target collection.
    \item Query the reference once per context, aggregate exposed probabilities through $g_j$, and place all residual probability mass in \texttt{OTHER}.
    \item Merge any declared category satisfying $M\pi_{ji}<c$ into \texttt{OTHER}; this map is fixed using the reference alone.
    \item Query each target route $M$ times with the exact same context and declared sampling settings.
    \item Map every returned string to one category.  Nonconforming text remains in \texttt{OTHER}; no target response is deleted.
    \item Compute \cref{eq:posterior-kl,eq:bias-corrected-kl}; save raw text, counts, reference categories, route statistics, null calibration, and request metadata.
    \item Simulate the joint route null, calculate one-sided route $p$ values, and apply Holm correction over the frozen matched-route family~\cite{holm1979simple}.
    \item If a route exposes top log probabilities, store them separately and compute a logprob-derived coarsened-KL comparator only as an auxiliary consistency check.
    \item Independently of the mean audit, collect complete long-sequence runs, rescore every retained target position with the trusted reference, and report the empirical run-level distribution without deleting extreme observations.
\end{enumerate}

\subsection{Outcome Mapping and Reference Support}
\label{app:implementation-details}

Each one-token task declares two to four exact digit labels.  An exact returned label enters its named category; whitespace-only, multi-token, missing, or otherwise nonconforming text enters \texttt{OTHER}.  The same map is applied to reference probability mass, target text counts, and the optional target-logprob comparator.  Consequently every target request contributes to the multinomial count.

The reference request exposes the top 20 token probabilities.  Probabilities of declared labels are retained when present, and all remaining probability is assigned to \texttt{OTHER}; the augmented vector is renormalized.  Reference-only pooling then merges categories whose expected count is below $c$.  Target output never changes this support.

\section{Experimental Protocol}
\label{app:repeated-context-protocol}

The protocol was frozen before formal collection.  It uses the requested DeepSeek V4 Flash 0731 revision, $J=12$ fixed contexts disjoint from five method-development contexts, $M=50$ target samples per context--route cell, matched temperature 1, a one-token output limit, $\tau=1$, $c=1$, 20,000 parametric-null draws, and analysis seed 20260814.  The official reference returns its top 20 log probabilities.  Pooling thresholds $c\in\{2,5\}$ are prespecified sensitivities.  The official V4 documentation establishes the model family~\cite{deepseek2026v4}.  Here ``0731'' denotes the requested revision label, not the collection date: the 2026-07-31 change log names \texttt{DeepSeek-V4-Flash-0731} and describes it as retaining the Preview architecture and size with new post-training~\cite{deepseek2026changelog}.

The formal audit contains seven matched routes, each evaluated on the same 12 contexts and reference vectors with $M=50$, temperature 1, and a one-token limit.  Target log probabilities are not requested from Aliyun, Ark, Baidu, or DeepInfra.

\begin{table}[ht]
\centering
\caption{Configuration fields preserved for the repeated-request audit.  The requested revision is DeepSeek V4 Flash 0731 for every row.  Optional target log probabilities are provider-controlled and never enter the text-count statistic.}
\label{tab:repeated-route-configuration}
\small
\begin{tabular}{lllc}
\toprule
Route snapshot & Access path & Precision status & Target logprobs \\
\midrule
Official self-check & Official API & not declared & yes \\
Aliyun 0731 & Aliyun Model Studio & not declared & no \\
Ark 0731 & Volcengine Ark & not declared & no \\
Baidu Qianfan 0731 & Baidu Qianfan & not declared & no \\
StreamLake & OpenRouter & not declared & yes \\
DeepInfra FP8 & OpenRouter & provider-declared FP8 & no \\
DigitalOcean & OpenRouter & not declared & yes \\
\bottomrule
\end{tabular}
\end{table}

The contexts are fair or weighted virtual choices over prompt-declared digit supports.  Exact prompt strings and supports are present in the released protocol and summary artifacts.  The formal matched design contains 4,200 one-token target calls and 12 shared reference calls.  The normalized artifacts do not preserve provider region, \texttt{top\_p}, stream mode, timeout, connection reuse, conversation or session identifiers, per-request seed support, randomized route and context interleaving, or retry lineage.  We therefore do not claim those controls were randomized or standardized, and route--time correlation, correlated routing, and unreconstructable retries remain threats to calibration and exact operational reproduction.

\subsection{Auxiliary Comparator and Sensitivity Inference}

The auxiliary comparison centers $\widehat K^{\mathrm{bc}}_{jr}$ and the logprob-derived comparator within each context.  It reports Pearson and Spearman association across the complete route--context grid with target log probabilities.  The exact one-sided test enumerates a common permutation of route-condition labels across all contexts, preserving each route condition as a repeated-measures block.  Route-level detection averages the 12 context estimates, simulates their joint null under independent multinomials from the fixed reference probabilities, and applies Holm correction across the seven matched routes.

The prior-free sensitivity statistic is the following order-two unbiased Pearson-divergence U-statistic~\cite{hoeffding1948ustatistics}:
\begin{equation}
 \widehat D_{\chi^2,jr}
 =\sum_i\frac{N_{jri}(N_{jri}-1)}{M(M-1)\pi_{ji}}-1,
 \qquad
 \mathbb E[\widehat D_{\chi^2,jr}]=D_{\chi^2}(\theta_{jr}\|\pi_j).
 \label{eq:chi-square-u}
\end{equation}
Expanding $u\log u$ around $u=1$ gives, for local alternatives, $D_{\mathrm{KL}}=\tfrac12D_{\chi^2}+O(\|\theta/\pi-1\|^3)$, so $\widehat D_{\chi^2}/2$ checks estimator dependence without a Bayesian prior.

Averaging the primary estimates first by route gives Pearson $r=0.989$ and perfect rank agreement with the logprob-derived comparator.  The prior-free statistic has context-centered Pearson $r=0.926$ and Spearman $\rho=0.560$.  Raising the minimum reference expected count from $c=1$ to $c=2$ and $c=5$ changes centered Pearson from $0.971$ to $0.942$ and $0.951$; the corresponding centered Spearman values are $0.657$, $0.653$, and $0.755$.  With three matched, logprob-capable route conditions, every exact one-sided common-route permutation test gives $p=1/6=0.167$ at the observed ordering.  This is descriptive agreement with a target-controlled field, not confirmatory validation against ground truth.

Four additional DigitalOcean batches were collected after the primary matched-route window.  They are labeled exploratory replications rather than part of the main route comparison.  They show the observed dispersion of $S_{r,b}$ over four additional windows; four batches are insufficient to estimate a stable cross-window mean $A_r$ or the latent excess maximum $X_r(m)$.

\subsection{Long-Sequence Run-Level Protocol}
\label{app:tail-protocol}

The run-level study uses 20 paired replicates for each of Official, Aliyun, Ark, Baidu Qianfan, StreamLake, DeepInfra FP8, and DigitalOcean.  Within a replicate, all seven routes receive the same freshly generated nonce challenge; across replicates, nonces are independent.  Each admitted target sequence contains all 500 prespecified scored positions.  A separate set of 20 official-control sequences uses disjoint nonces, producing 160 complete sequences and 80,000 scored positions in total.  The Official row in \cref{tab:tail-results} uses the 20 paired Official runs.  The disjoint Official controls are retained as a diagnostic partition in the released run-level data and do not enter the reported route table, figures, or route comparisons.

For every target position, the trusted reference returns the conditional probability distribution under the exact target prefix.  The analysis stores observed surprise, reference entropy, information variance, and the centered residual in \cref{eq:centered-surprise}.  It computes $D_{r,b}(n)$ at the prespecified position budgets
\begin{quote}\small
$n\in\{10,20,30,40,50,75,100,125,150,175,200,250,300,350,400,450,500\}$.
\end{quote}
The complete sequence is the statistical unit.  Pointwise 95\% intervals resample the 20 runs as clusters with 20,000 bootstrap draws; the same sampled run indices are reused across all $n$ to preserve paired convergence differences.  No control-derived cutoff or binary stability label is constructed.

The primary run-level report uses all 20 values of $D_{r,b}(500)$.  It reports the median, standard deviation, empirical quantiles, and maximum.  The empirical $q_{0.95}$ uses NumPy 1.26.4's default \texttt{method=linear}, corresponding to the commonly used type-7 sample quantile~\cite{hyndman1996sample,harris2020array}.  The maximum is an observed sample statistic rather than an estimate of the population endpoint, and none of these summaries estimates $X_r(m)$.

The released analysis environment is recorded in \path{requirements-analysis.txt}: Python 3.11.6, NumPy 1.26.4, SciPy 1.17.1, and Matplotlib 3.11.1.  SciPy supplies the reported correlations, Mann--Whitney test, and central and noncentral chi-square functions~\cite{virtanen2020scipy}; bootstrap resampling is implemented directly with NumPy.

\subsection{Downstream Evaluation Protocol}
\label{app:downstream-protocol}

GPQA-Diamond uses all 198 items in the benchmark's highest-quality subset~\cite{rein2024gpqa,rein2026gpqadataset} for each of the seven routes in \cref{tab:tail-results}.  The frozen item order and answer parser are shared across routes.  Reported route accuracies use Wilson 95\% intervals~\cite{wilson1927probable}; Pearson and Spearman significance enumerates all $7!$ route-label permutations.

Terminal-Bench 2.1 uses all 89 tasks per route.  We cite both the Terminal-Bench paper, which introduces the 89-task 2.0 benchmark, and the official 2.1 dataset used here~\cite{merrill2026terminalbench,terminalbench2026dataset}.  Runs use the Terminus-2 agent distributed with Harbor 0.21.0~\cite{harbor2026v021}.  The benchmark was collected from 2026-08-13 10:09 to 2026-08-15 02:51 China Standard Time and is an exploratory observational study, not a time-aligned continuation of the audit windows.  Confirmed provider-side \texttt{APIError} trials for Ark and Qianfan are replaced only by targeted same-task reruns.  Final \texttt{AgentTimeoutError}, \texttt{VerifierTimeoutError}, and runtime-error outcomes remain zero-reward observations.  DigitalOcean was collected as a separate replication under the same task protocol.  The task-exposure proxy is the median model-request count for the same task across the five non-DigitalOcean matched routes and uses no DigitalOcean-derived complexity measure.  Quartile uncertainty resamples whole tasks with 20,000 bootstrap draws~\cite{efron1993bootstrap}.

\section{Downstream Benchmark Results}
\label{app:downstream-results}

All seven routes answer every GPQA-Diamond item, and all scored responses parse successfully.  Accuracy spans 69.7--74.7\% with overlapping Wilson intervals.  Median $D_{500}$ has Pearson $r=0.347$ (exact $p=0.446$) and Spearman $\rho=0.464$ ($p=0.302$) with accuracy, providing no evidence of a negative route-level association.

\begin{figure}[ht]
    \centering
    \includegraphics[width=0.72\textwidth]{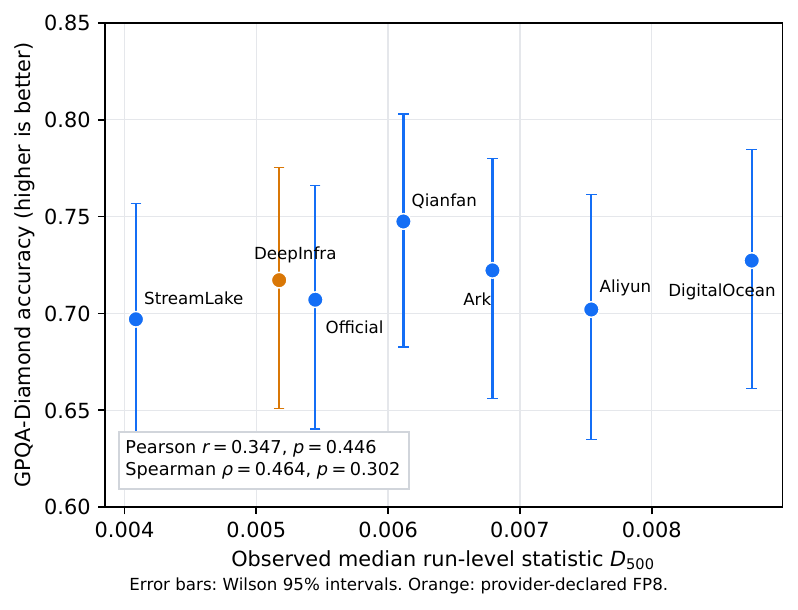}
    \caption{GPQA-Diamond accuracy versus median run-level $D_{500}$.  Error bars are Wilson 95\% intervals; the route-level association is not statistically supported.}
    \label{fig:gpqa}
\end{figure}

Terminal-Bench produces one terminal outcome for every task and route.  Only confirmed provider-side \texttt{APIError} trials are replaced by targeted same-task reruns: 57 for Ark and 14 for Qianfan.  All final agent, verifier, and runtime failures remain zero-reward outcomes.  DigitalOcean is collected as a separate replication under the same task protocol.

\begin{table}[ht]
\centering
\caption{Complete Terminal-Bench 2.1 route results after targeted API-error replacement.  Every denominator is 89 tasks.}
\label{tab:terminal-bench-results}
\small
\begin{tabular}{lrrrr}
\toprule
Route & $S_r$ & Passes & Pass rate $\uparrow$ & Errors $\downarrow$ \\
\midrule
Official & $-0.0007$ & 48 & 53.9\% & 22 \\
Aliyun 0731 & $0.5704$ & 45 & 50.6\% & 26 \\
Ark 0731 & $0.1591$ & \textbf{53} & \textbf{59.6\%} & \textbf{14} \\
Baidu Qianfan 0731 & $0.1875$ & 38 & 42.7\% & 32 \\
StreamLake & $0.2950$ & 52 & 58.4\% & 22 \\
DeepInfra FP8 & $0.1185$ & 46 & 51.7\% & 23 \\
DigitalOcean & $0.1250$ & 33 & 37.1\% & 48 \\
\bottomrule
\end{tabular}
\end{table}

Across the seven routes, $S_r$ has Pearson $r=0.081$ (exact $p=0.873$) and Spearman $\rho=-0.036$ ($p=0.963$) with Terminal-Bench pass rate.  The within-window mean statistic alone therefore does not order Terminal-Bench performance.

For the exposure analysis, tasks are sorted by the median model-request count for the same task across five matched non-DigitalOcean routes.  This construction contains no DigitalOcean-derived complexity measure.  All 47 DigitalOcean \texttt{AgentTimeoutError} outcomes remain failures because removing them would condition on the observed outcome.

\begin{table}[ht]
\centering
\caption{DigitalOcean success across task-exposure quartiles.  The comparator is the same-task mean over the five matched routes used to construct exposure.}
\label{tab:do-horizon}
\small
\begin{tabular}{lrrrr}
\toprule
Exposure quartile & Tasks & DO pass rate & Comparator pass rate & DO agent timeouts \\
\midrule
Q1 (shortest) & 23 & 82.6\% & 76.5\% & 4 \\
Q2 & 22 & 31.8\% & 62.7\% & 10 \\
Q3 & 22 & 18.2\% & 37.3\% & 16 \\
Q4 (longest) & 22 & 13.6\% & 28.2\% & 17 \\
\bottomrule
\end{tabular}
\end{table}

DigitalOcean's comparator-relative gaps are $+6.1$, $-30.9$, $-19.1$, and $-14.6$ percentage points from Q1 to Q4, so the deficit is not monotonic across quartiles.  The prespecified endpoint contrast shows that the Q4 gap is 20.6 percentage points worse than the Q1 gap; the task-bootstrap 95\% interval is $[-36.0,-4.5]$ percentage points and the one-sided nonnegative-tail probability is $0.0059$.  Failed DigitalOcean tasks also have larger exposure than successful tasks (median 36.5 versus 11 requests; one-sided Mann--Whitney $p=3.87\times10^{-7}$)~\cite{mann1947test}.  However, 47 of its 56 non-pass outcomes (84\%) are \texttt{AgentTimeoutError} outcomes.  The serving deviations reflected by Ventor-QTest's empirical upper-tail statistic may have contributed to these failures, but the lack of time-aligned probe and benchmark windows prevents confirmed causal attribution.

\section{Fixed-sample Request Planning}
\label{app:repeated-power}

A finite request budget is defined relative to a minimum effect of interest.  For the frozen contexts, a route's within-window mean coarsened KL is
\begin{equation}
 \delta=J^{-1}\sum_jD_{\mathrm{KL}}(\theta_{jr}\|\pi_j).
 \label{eq:average-coarse-kl}
\end{equation}
For acquisition planning, the local asymptotic multinomial likelihood-ratio model~\cite{vandervaart1998asymptotic} gives
\begin{equation}
 \nu(M)=\sum_{j=1}^{J}\{k_j(M)-1\},\qquad
 G_r^2\ \dot\sim\ \chi^2_{\nu(M)}(\lambda),\qquad
 \lambda\simeq 2JM\delta,
 \label{eq:repeated-power-noncentrality}
\end{equation}
where $k_j(M)$ is the number of categories remaining after reference-only pooling.  For $R$ simultaneous route comparisons, planning uses the conservative threshold $\alpha_*=0.05/R$.  Approximate power is
\begin{equation}
 \mathcal P(M,\delta)
 =1-F_{\chi^2_{\nu(M)}(2JM\delta)}
 \!\left(\chi^2_{\nu(M),1-\alpha_*}\right).
 \label{eq:repeated-power}
\end{equation}
Integer $M$ is enumerated after applying the same $M\pi_{ji}\ge1$ pooling rule used by the estimator, so support changes are included in $\nu(M)$.

\begin{table}[ht]
\centering
\caption{Fixed-sample planning for $J=12$, 90\% power, and family-wise level 0.05 across $R=6$ route comparisons.  Calls are one-token target requests per route; 12 reference distributions are shared across routes.}
\label{tab:repeated-power}
\small
\begin{tabular}{rrrr}
\toprule
Minimum mean coarsened KL, $\delta$ & Repeats/context, $M$ & Target calls/route & $\nu(M)$ \\
\midrule
$0.010$ & 159 & 1,908 & 26 \\
$0.020$ & 79  & 948   & 25 \\
$0.030$ & 52  & 624   & 24 \\
$0.050$ & 30  & 360   & 21 \\
$0.100$ & 15  & 180   & 19 \\
\bottomrule
\end{tabular}
\end{table}

At $M=50$, $\nu=24$, and inversion of \cref{eq:repeated-power} gives a 90\% planning effect floor of $\delta=0.0308$.  The weakest observed third-party score is $0.1185$, although the comparison is descriptive because the effect is estimated from the audit data.  Equations~\eqref{eq:repeated-power-noncentrality}--\eqref{eq:repeated-power} are local-asymptotic planning approximations; actual route claims use the 20,000-draw protocol-specific null.

\section{Mean-Component Cost Accounting}
\label{app:cost-accounting}

The following calculation covers only the repeated-request mean component.  The 12 exact contexts were metered once against official DeepSeek V4 Flash with one output token per request.  One complete pass uses 432 input and 12 output tokens.  At the rates captured on 2026-08-14 in the released artifact \path{data/repeated_context_cost.json}, cache-miss input cost \$0.14 and output cost \$0.28 per million tokens.  The artifact records the then-current official pricing page~\cite{deepseek2026pricing}; because that page is mutable, the artifact rather than its present contents is the evidence for these historical rates.  All input is conservatively charged as a cache miss.  A 600-call target route uses 21,600 input and 600 output tokens; the shared reference pass adds 432 input and 12 output tokens.  The single-route list-price equivalent is
\begin{equation}
 C_1=\frac{(21{,}600+432)(0.14)+(600+12)(0.28)}{10^6}
 =\text{\$}0.00325584\simeq\text{\$}0.0033.
 \label{eq:audit-price}
\end{equation}
The seven-route matched audit costs \$0.0224 under these assumptions.  Cache-hit discounts, provider-specific pricing, retries, taxes, and minimum charges are excluded.

\section{Released Experimental Data}
\label{app:data}

The release includes the frozen protocols, categorized counts for 4,200 matched-route responses, 12 reference category distributions, context-level estimates, route-level joint-null tests, and pooling sensitivities.  It also includes all 20 complete $D_{500}$ values for each of seven paired routes, convergence summaries, complete 198-item GPQA results, all 89 Terminal-Bench task outcomes per route, and the task-exposure audit.  The JSON and CSV artifacts reproduce \cref{fig:repeated-context-validation,fig:gpqa,tab:repeated-context-results,tab:tail-results,tab:terminal-bench-results,tab:do-horizon,tab:repeated-power}; authorization headers and API keys are excluded.

\end{document}